\documentclass[Physsubmission, Phys]{SciPost}

\hypersetup{
    colorlinks,
    linkcolor={red!50!black},
    citecolor={blue!50!black},
    urlcolor={blue!80!black}
}

\usepackage[bitstream-charter]{mathdesign}
\DeclareSymbolFont{usualmathcal}{OMS}{cmsy}{m}{n}
\DeclareSymbolFontAlphabet{\mathcal}{usualmathcal}

\begin{document}

\begin{center}{\Large \textbf{Primary Cosmic Ray Energy Spectrum and Mean 
Mass Composition Using Data from the TAIGA Astrophysical Complex\\
}}\end{center}

\begin{center}
V. Prosin\textsuperscript{1$\star$}
I. Astapov\textsuperscript{6}
P. Bezyazeekov\textsuperscript{2}
E. Bonvech\textsuperscript{2}
A. Borodin\textsuperscript{7}
A. Bulan\textsuperscript{1}
A. Chiavassa\textsuperscript{4}
D. Chernov\textsuperscript{1}
A. Dyachok\textsuperscript{2}
A. Gafarov\textsuperscript{2}
A. Garmash\textsuperscript{11,9}
V. Grebenyuk\textsuperscript{7,8}
O. Gress\textsuperscript{2}
E. Gress\textsuperscript{2}
T. Gress\textsuperscript{2}
A. Grinyuk\textsuperscript{7}
O. Grishin\textsuperscript{2}
A. D. Ivanova\textsuperscript{2}
A. L. Ivanova\textsuperscript{9,2}
N. Kalmykov\textsuperscript{1}
V. Kindin\textsuperscript{6}
S. Kiryuhin\textsuperscript{2}
R. Kokoulin\textsuperscript{6}
K. Komponiets\textsuperscript{6}
E. Korosteleva\textsuperscript{1}
V. Kozhin\textsuperscript{1}
E. Kravchenko\textsuperscript{9,11}
A. Kryukov\textsuperscript{1}
L. Kuzmichev\textsuperscript{1}
A. Lagutin\textsuperscript{10}
M. Lavrova\textsuperscript{7}
Y. Lemeshev\textsuperscript{2}
B. Lubsandorzhiev\textsuperscript{3}
N. Lubsandorzhiev\textsuperscript{1}
A. Lukanov\textsuperscript{3}
D. Lukyantsev\textsuperscript{2}
S. Malakhov\textsuperscript{2}
R. Mirgazov\textsuperscript{2}
R. Monkhoev\textsuperscript{2}
E. Okuneva\textsuperscript{6}
E. Osipova\textsuperscript{1}
A. Pakhorukov\textsuperscript{2}
A. Pan\textsuperscript{7}
L. Panasenko\textsuperscript{11}
L. Pankov\textsuperscript{2}
A. D. Panov\textsuperscript{1}
A. Petrukhin\textsuperscript{6}
I. Poddubny\textsuperscript{2}
D. Podgrudkov\textsuperscript{1}
V. Poleschuk\textsuperscript{2}
V. Ponomareva\textsuperscript{2}
E. Popova\textsuperscript{1}
E. Postnikov\textsuperscript{1}
V. Ptuskin\textsuperscript{5}
A. Pushnin\textsuperscript{2}
R. Raikin\textsuperscript{10}
A. Razumov\textsuperscript{1}
G. Rubtsov\textsuperscript{3}
E. Ryabov\textsuperscript{2}
Y. Sagan\textsuperscript{7,8}
V. Samoliga\textsuperscript{2}
A. Silaev\textsuperscript{1}
A. Silaev(junior)\textsuperscript{1}
A. Sidorenkov\textsuperscript{3}
A. Skurikhin\textsuperscript{1}
A. Sokolov\textsuperscript{9,11}
L. Sveshnikova\textsuperscript{1}
V. Tabolenko\textsuperscript{2}
A. Tanaev\textsuperscript{2}
B. Tarashchansky\textsuperscript{2}
M. Y. Ternovoy\textsuperscript{2}
L. Tkachev\textsuperscript{7}
R. Togoo\textsuperscript{12}
N. Ushakov\textsuperscript{3}
A. Vaidyanathan\textsuperscript{11}
P. Volchugov\textsuperscript{1}
N. Volkov\textsuperscript{10}
D. Voronin\textsuperscript{3}
A. Zagorodnikov\textsuperscript{2}
A. Zhaglova\textsuperscript{2}
D. Zhurov\textsuperscript{2,13}
I. Yashin\textsuperscript{6}
\end{center}

\begin{center}

{\bf 1} Skobeltsyn Institute of Nuclear Physics, Moscow State University, Moscow, Russia
\\
{\bf 2} Applied Physics Institute of Irkutsk State University, Irkutsk, Russia
\\
{\bf 3} Institute for Nuclear Research of the RAS, Troytsk, Moscow, Russia
\\
{\bf 4} Dipartimento di Fisica Generale Universiteta di Torino and INFN, Turin, Italy
\\
{\bf 5} Puskov Institute of Terrestrial Magnetism, Ionosphere and Radio Wave Propagation of the RAS, Troitsk, Moscow. Russia
\\
{\bf 6} National Research Nuclear University MEPhI, Moscow, Russia
\\
{\bf 7} Joint Institute for Nuclear Research, Dubna, Moscow Region, Russia
\\
{\bf 8} DUBNA University, Dubna, Moscow Region, Russia
\\
{\bf 9} Budker Institute of Nuclear Physics SB RAS, Novosibirsk, Russia
\\
{\bf 10} Altai State Univeristy, Barnaul, Russia
\\
{\bf 11} Novosibirsk State University, Novosibirsk, Russia
\\
{\bf 12} Institute of Physics and Technology Mongolian Academy of Sciences, Ulaanbaatar, Mongolia
\\
{\bf 13} Irkutsk National Research Technical University, Irkutsk, Russia
\\

* v-prosin@yandex.ru
\end{center}
 
\begin{center}
\today
\end{center}


\definecolor{palegray}{gray}{0.95}
\begin{center}
\colorbox{palegray}{
  \begin{tabular}{rr}
  \begin{minipage}{0.1\textwidth}
    \includegraphics[width=30mm]{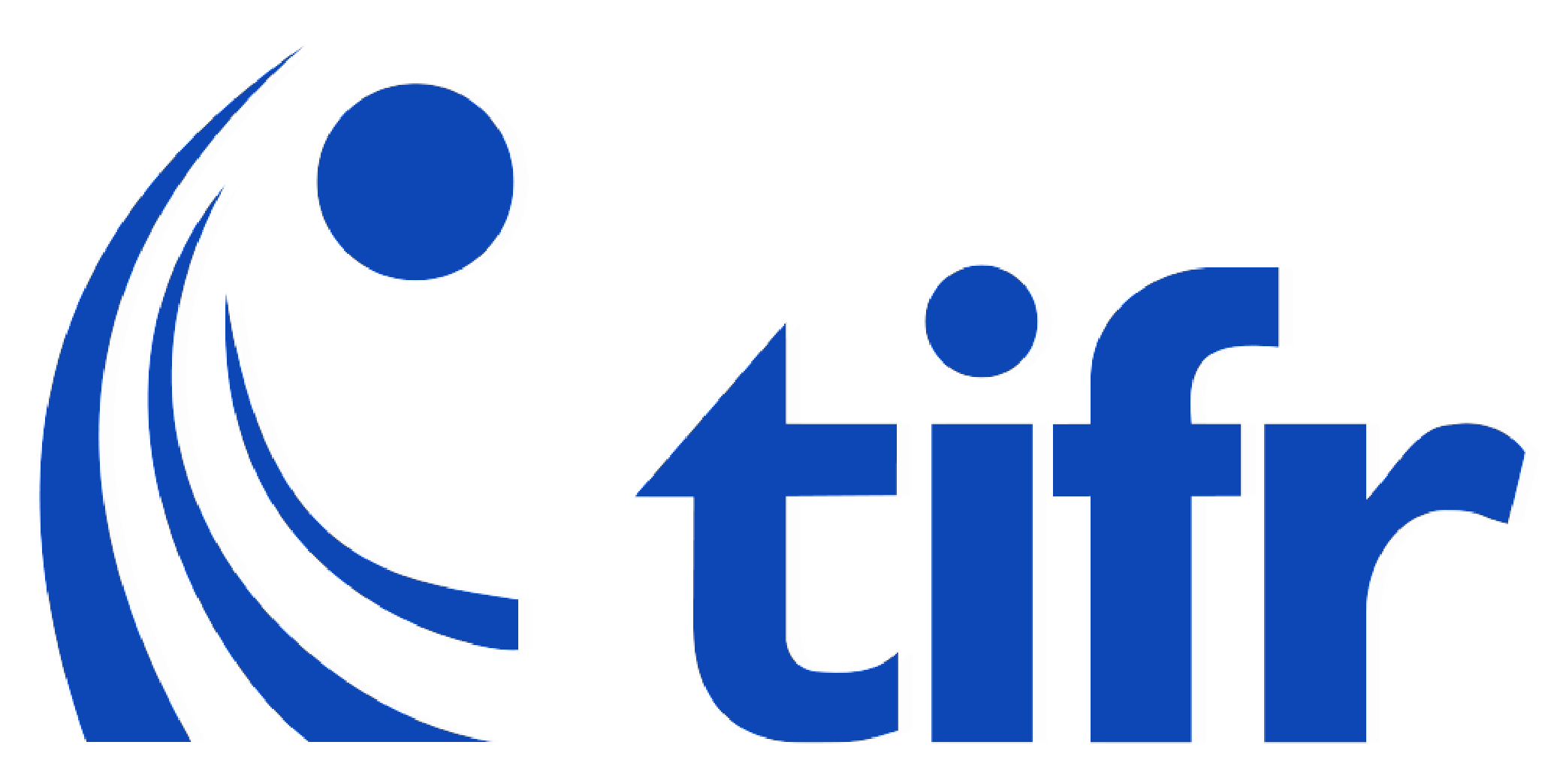}
  \end{minipage}
  &
  \begin{minipage}{0.85\textwidth}
    \begin{center}
    {\it 21st International Symposium on Very High Energy Cosmic Ray Interactions (ISVHECRI 2022)}\\
    {\it Online, 23-27 May 2022}\\
    \doi{10.21468/SciPostPhysProc.?}\\
    \end{center}
  \end{minipage}
\end{tabular}
}
\end{center}

\section*{Abstract}
{\bf
The
corrected dependence of the mean depth of the EAS maximum $X_{max}$ on 
the energy was obtained 
from the data of the Tunka-133 array for 7 years and the TAIGA-HiSCORE 
array for 2 years. The parameter $\langle\ln A\rangle$, characterizing the mean
mass compositon was derived from these results. 
The differential energy spectrum of primary cosmic rays in the energy range of 
$2\cdot 10^{14}$ – $2\cdot 10^{16}$\,eV was reconstructed using the new parameter
$Q_{100}$ the Cherenkov light flux at the core distance 100 m. Change of the parameter 
for the energy reconstuction for the TAIGA-HiSCORE from $Q_{200}$ to $Q_{100}$ provides 
a decreasing energy threshold for the spectrum to about 200 TeV.}

\vspace{10pt}
\noindent\rule{\textwidth}{1pt}
\tableofcontents\thispagestyle{fancy}
\noindent\rule{\textwidth}{1pt}
\vspace{10pt}

\section{Introduction}
\label{sec:intro}
The energy spectrum and mass composition of primary
cosmic rays are the main characteristics that can
be obtained by studying extensive air showers
(EAS). The total flux of Cherenkov light is proportional
to the total energy scattered by the shower in the
atmosphere. The lateral distribution function (LDF) of the EAS Cherenkov light
reflects the position of the shower development maximum, 
which in turn characterizes the mass of
the primary particle. The main aim of the present work is to achieve
a lower energy threshold by the new method of energy reconstruction, 
descibed below.

\section{Brief description of the arrays}
Several arrays that detect the EAS Cherenkov
light were successively constructed in the Tunka
Valley. The
most productive arrays were the Tunka-25 (2000 - 2005) \cite{t25}, 
which consisted of 25 detectors with a
sensitive area of 0.1 $m^2$ each, covering a total area of
approximately 0.1 $km^2$, and the Tunka-133 (2009 - 2017)
\cite{t133}, consisted finally of 175 detectors
with a sensitive area of 0.03 $m^2$, covering an area of
approximately 3 $km^2$.
The experimetal data was accumulated
over 350 clear moonless nights. The total time of the
data acquisition was 2175h.
Their modern successor is the TAIGA-HiSCORE
\cite{his} array, a part of the TAIGA experimental complex
\cite{taiga}.
TAIGA-HiSCORE single
station has a sensitive area of 0.5 $m^2$.
Every station has it's own trigger. The stations are merged to an EAS 
event in case of $\geq 3$  station hits coincident inside a time gate of 3 $\mu$s.  
This work presents the TAIGA-HiSCORE data
that was obtained using 67 stations (two first clusters) during
135 clear moonless nights in the seasons of 2019 - 2020 and 2020 - 2021. 
The total data acquisition time was 327 h. 

\begin{figure} 
\centering
  \begin{tabular}{rr}
  \begin{minipage}{0.5\textwidth}
    \includegraphics[width=70mm]{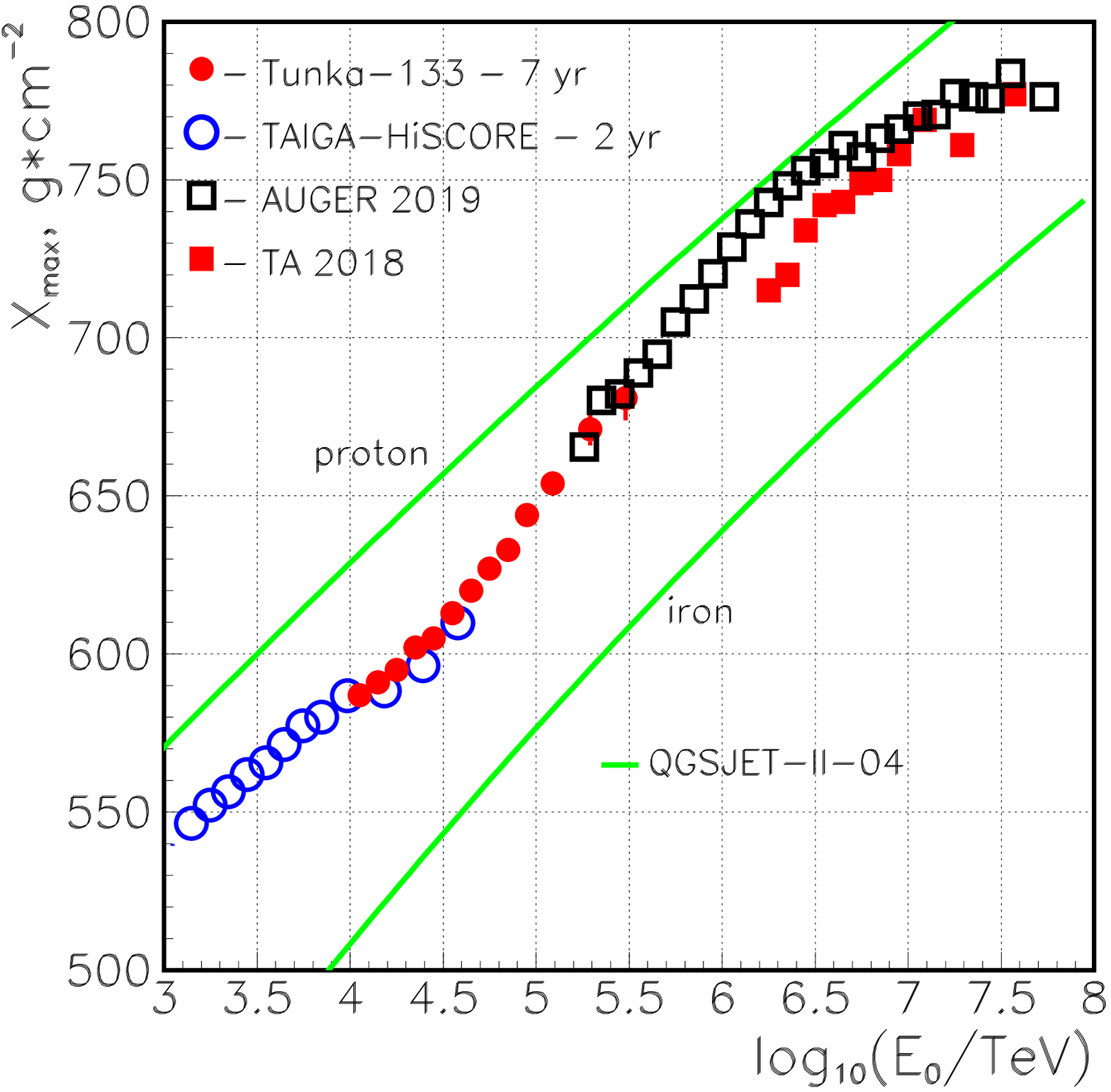}
\caption{Mean experimental $X_{max}$}
  \end{minipage}
  &
  \begin{minipage}{0.5\textwidth} 
   \includegraphics[width=70mm]{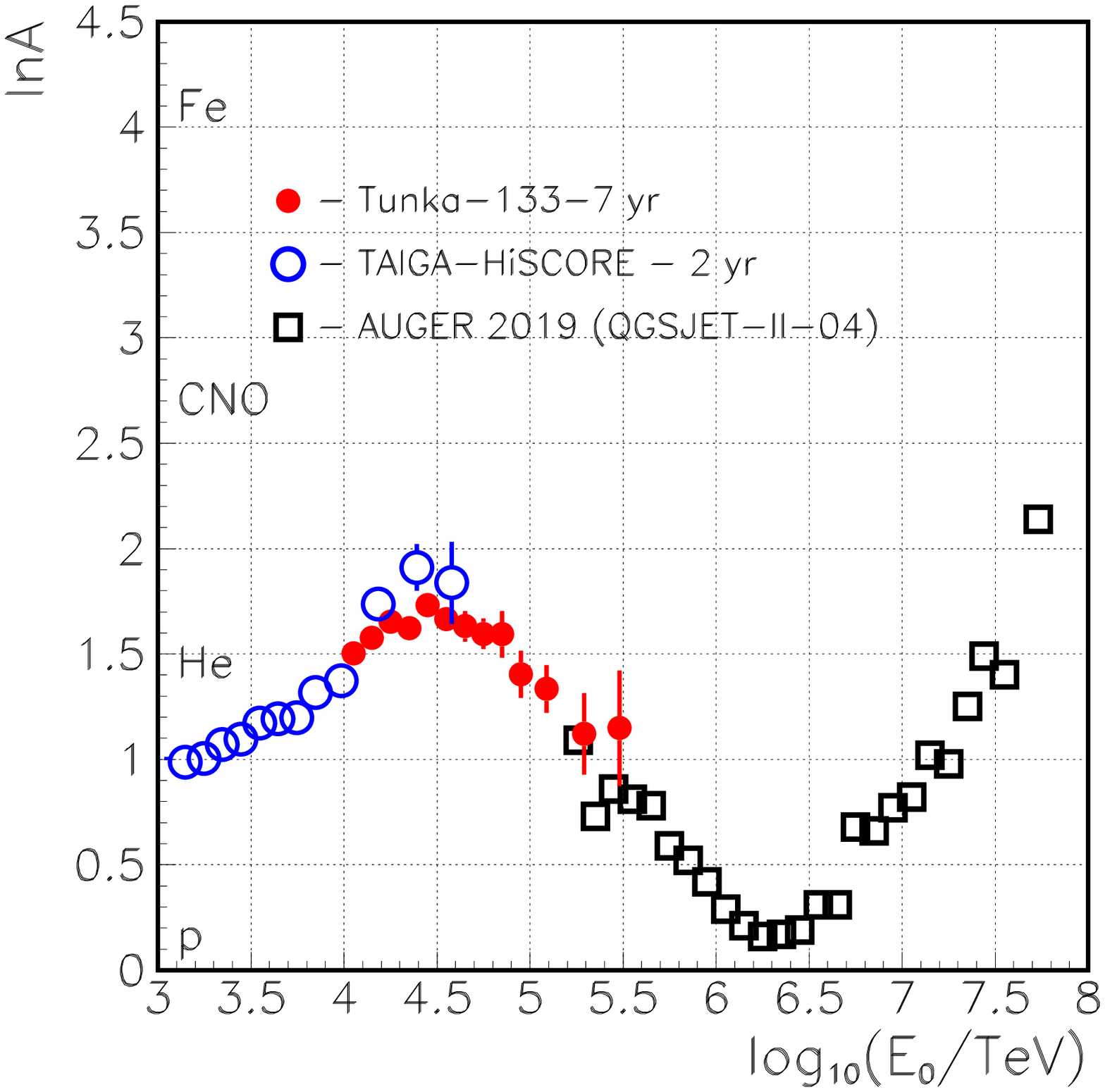}
\caption{Mean $\langle\ln A\rangle$}
\end{minipage}
\end{tabular}
\end{figure}

\section{Reconstruction of the EAS parameters}
The reconstruction of the EAS parameters for the Tunka-133 array is
described in \cite{t133}. The same algorithms and the fitting functions
are used for the TAIGA-HiSCORE data processing \cite{his}. 
We use the ratio
$P = Q(80)/Q(200)$ as a quantitative parameter of LDF steepness. 
Here, $Q(R)$ is the Cherenkov light flux at a distance, R, in meters.
One has to control that there are measurements of light flux at 
core distances $R_c \geq 200$\,m and $R_c \geq 80$\,m. 
The first of these conditions is applied to the events
for the primary energy $E_0 \geq 10^{16}$\,eV for Tunka-133 and 
$E_0 \geq 10^{15}$\,eV for the TAIGA-HiSCORE.
CORSIKA simulation \cite{xms} confirmed that
the Cherenkov light LDF steepness is determined solely by the thickness of the
atmosphere between the array and the
depth of the EAS maximum ($\Delta X_{max} = X_0/sec\theta - X_{max}$). Here, $X_0$ is
the total depth of the atmosphere.
The calculated connection between $P$ and $\Delta X_{max}$, 
inside the limited range of 
parameter $P$ from 2.5 to 9, can be fitted inside the 
following expression \cite{xms}:

\begin{equation}
\Delta X_{max}=\left\{
\begin{array}{lc}
\!\ 929 - 103\cdot P, & \mbox{if}\hspace{5mm} P \leq 3.9\\
\!\ 882 - 91\cdot P, & \mbox{if}\hspace{5mm} P > 3.9\\
\end{array} 
\right.
\end{equation}

 The standard deviation of simulated
points from the fitting line for this range is approximately
15 $g/cm^2$ \cite{xms}.

\section{Mean experimental depth of the EAS maximum}
The above described parameter of the LDF steepness $P$ was
applied to analyze the data of both the Tunka-133 and
TAIGA-HiSCORE arrays. The depth of the maximum is found by the formula:

\begin{equation}
X_{max} = 965/sec\theta - \Delta X_{max}			
\end{equation}

where 965 $g/cm^2$ is the total depth of the atmosphere
at the location of the arrays in the Tunka Valley.
To obtain undistorted estimations
of the depth of the maximum, showers are
selected for the zenith angle $\theta \leq 30^{\circ}$ and the energy
above $10^{16}$\,eV for the Tunka-133 array and
above $10^{15}$\,eV for TAIGA-HiSCORE. We have 69000 events for 
7 years of operation of the Tunka-133 and 380000 events for 2 years of 
operation of TAIGA-HiSCORE. 
The expermental results
are shown in Fig.\,1. The data of both
arrays, despite the difference in their geometry, agree
well with each other, providing a wide energy range
from $10^{15}$ to $3\cdot 10^{17}$\,eV. Our experimental data are
compared with the direct measurements of the depth
of the maximum obtained by observing the fluorescent
EAS light at the Pierre Auger Observatoy (PAO) \cite{pao} and Telescope Array
(TA) \cite{ta}. A close agreement of our data with the PAO
data is observed at an energy of $\sim 3\cdot 10^{17}$\,eV.
All the experimental results are compared with theoretical
curves calculated using the QGSJET-II-04
model \cite{ost} for primary protons and iron nuclei.

Fig.\,2
shows the results of recalculation from the mean depth
of the maximum to the parameter $\langle\ln A\rangle$ the 
average logarithm of the atomic number
using the QGSJET-II-04 model. 
Qualitatively, the
behavior of the mean mass composition repeats what
was published in our previous studies \cite{xmold} 
becoming heavier in the energy range of 
$3\cdot 10^{15}$ - $3\cdot 10^{16}$\,eV 
and lighter with a further
increase in energy. However, the mean composition in
the entire energy range under consideration is estimated
as mostly light. 

\section{Cherenkov light flux at a core distance 100 m as a new estimator of
energy}  
The TAIGA-HiSCORE array structure is a square net of stations with a step 
of about 100 m. So the minimal core distance for which light flux can be 
reconstructed for almost all the events is about 100 m. Our previous parameter 
for the energy reconstruction 
was light flux at a core distance 200 m ($Q_{200}$) \cite{his}. 
When the EAS zenith angle $\theta$ changes 
from $0^{\circ}$ to $45^{\circ}$, $Q_{200}$ changes by less than 10\%. Therefore,
it was assumed in \cite{his}, that $Q_{200}$ does not depend on $\theta$ for a fixed 
particle energy. 
It was found by the new CORSIKA simulation that light flux $Q_{100}$ depends
on the zenith angle $\theta$ sigificantly more, changing by about 2.5 times with the same 
change in $\theta$. So first one needs to recalculate 
from the measured light flux to the $\theta = 0^{\circ}$ using the new CORSIKA 
results:

\begin{equation}
\log_{10}(Q_{100}(0)) = \log_{10}(Q_{100}(\theta)) +
(\sec \theta - 1)\cdot(1.25 - 0.083\cdot \log_{10}(Q_{100}(\theta))
\end{equation}

Then $Q_{100}(0)$ can be recalculated to the primary energy $E_0$ using 
the result of the new CORSIKA simulation:

\begin{equation}
\log_{10}(E_0/GeV) = 0.88\cdot \log_{10}(Q_{100}(0)) + 5.14
\end{equation}

\begin{figure}[ht]
\centering
\includegraphics[width=0.5\textwidth]{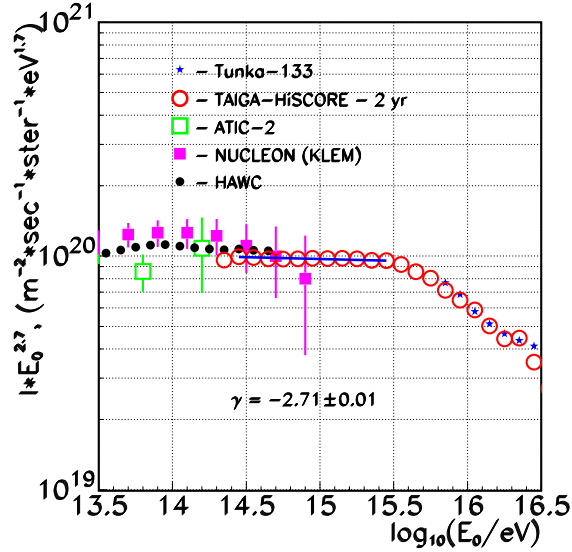}
\caption{Differential primary CR spectrum by TAIGA-HiSCORE data}
\label{fig:3}
\end{figure}

\section{Experimental energy spectrum by the data of TAIGA-HiSCORE}
The experimental energy estimation differs from that described at the previous 
section 
because the real atmosphere light absorbtion is different from night to night 
in contradiction with standard absorption assumed in CORSIKA simulations. 
So first we obtain the integral energy spectrum for the single night using the 
expression (4). Then we normalize this spectrum to the reference energy 
spectrum measured by the QUEST 
experiment \cite{kor}. The mean difference of normalization constant 
from that in expression (4) is 0.03.
The differential energy spectrum obtained from the data of TAIGA-HiSCORE 
array is shown in Fig.\,3. Efficiency of the events at the 
first left point (starting from the energy $2\cdot 10^{14}$\,eV) is more than 95\%. 
Points for the lower energy obtained from the events with lower efficiency are removed.
The low energy points of our spectrum are in 
good agreement with direct balloon \cite{atic}, satellite \cite{nucl}
and mountain \cite{hawc} measurements.

\section{Conclusions} 
The new estimations of $X_{max}$, derived from the 
steepness parameter $P = Q(80)/Q(200)$ provides good 
agreement between both the results of our arrays 
Tunka-133, TAIGA-HiSCORE and Tunka-133 and 
the direct measurements of $X_{max}$ at the 
Pierre Auger Observatory (PAO).

The primary composition, derived from $X_{max}$ is 
lighter than it seemed in our previous publications. 
It is mostly light (p+He) over the whole energy range.

The observed increase of $\langle \ln A\rangle$ in the energy 
range $10^{16}$ - $10^{17}$\,eV demands a new theoretical explanation.

Change of the parameter for the energy reconstruction for 
the TAIGA-HiSCORE from $Q_{200}$ to $Q_{100}$ provides 
a decreasing energy threshold for the spectrum to about 200 TeV.

The all particle energy spectrum over the energy range 200 TeV - 3 PeV 
follows a pure power law with index 
$2.71\pm 0.01$.

\section*{Acknowledgements}
The work was performed at the UNU “Astrophysical Complex of MSU-ISU” 
(agreement EB 075-15-2021-675). The work is supported by RFBR 
(grants 19-52-44002, 19-32-60003), the RSF(grants 19-72-20067(Section 2)), 
the Russian Federation Ministry of Science and High Education 
(projects FZZE-2020-0017, FZZE-2020-0024, and FSUS-2020-0039).

\nolinenumbers
\end{document}